\documentclass[A4paper, 11pt]{article}
\usepackage{amsmath,amssymb,bm,epsf,epsfig,graphicx}
\input epsf.sty

\usepackage{parskip}
\usepackage[utf8]{inputenc}
\usepackage[english]{babel}
\usepackage{amsmath}
\usepackage{latexsym}
\usepackage{amsfonts}
\usepackage{mathtools}
\usepackage{amssymb}
\usepackage{scalerel}
\usepackage{extpfeil}
\usepackage[left=3cm,right=3cm,top=3.1cm,bottom=3.1cm]{geometry}
\usepackage{cite}
\usepackage{tensor}
\usepackage{tikz}
\usepackage{tikz-cd}
\usepackage[mathscr]{euscript}
\usetikzlibrary{arrows.meta, bending}
\tikzset{C/.style={circle, minimum size=8mm,
		node contents={},
		append after command={\pgfextra{%
				\draw[-{Straight Barb[flex']}](\tikzlastnode.150) arc (450:110:2.8mm);}
	}}
}

\usepackage{hyperref}
\numberwithin{equation}{section}

\def\ket#1{|#1 \rangle}
\def\aver#1{\left\langle\, #1 \,\right\rangle}

\def \be {\begin{eqnarray}}
\def \ee {\end{eqnarray}}
\def \bal {\begin{align}}
\def \eal {\end{align}}
\def \bdm {\begin{displaymath}}
\def \edm {\end{displaymath}}

\def\0{\nonumber}

\let\realcite\cite
\renewcommand*{\cite}[1]{{\footnotesize\realcite{#1}}}

\begin{document}
	\begingroup\allowdisplaybreaks

\vspace*{1.1cm}

\centerline{\Large \bf String Field Theory}\vspace{.3cm}


\begin{center}

{\large Carlo Maccaferri\footnote{Email: maccafer at gmail.com}  }
\vskip .7 cm
{\it Dipartimento di Fisica, Universit\`a di Torino, \\ and INFN  Sezione di Torino \\
Via Pietro Giuria 1, I-10125 Torino, Italy}
\end{center}
\vspace*{9.0ex}

\centerline{\bf Abstract}
\bigskip
String Field Theory is a formulation of String Theory as a Quantum Field Theory in target space. It allows to tame the infrared divergences of String Theory and to approach its non-perturbative structure and background independence. This article  gives a concise overview on the subject and of some of the main recent  progress.
\vspace*{6.0ex}

Invited contribution for the {\it Oxford Research Encyclopedia of Physics} by Oxford University Press.


\newpage
\tableofcontents

\newpage

\section{Introduction and summary}
After more than fifty years from the Veneziano amplitude, the fundamental formulation of String Theory (ST) remains elusive. 
On the one hand there is the world-sheet formulation which is truly microscopic but which is only valid for infinitesimal string coupling and is highly background dependent. On the other hand there are other  non-perturbative background independent approaches such as Supergravity or other quantum theories of (super) Yang-Mills type which however necessarily miss some of the features of the extended nature of the string (although they can be in some cases holographically equivalent to a string theory).

In String Field Theory (SFT)  it is possible to keep an exact microscopic world-sheet description together with a complete space-time framework which follows the rules of Quantum Field Theory (QFT) and where non-perturbative contributions can be, at least in principle, coherently accounted for.

In few words, SFT is a formulation of ST as a QFT for an infinite number of fields (the various oscillation modes of the string)  in spacetime. This formulation allows to better treat some of the shortcomings of the usual on-shell formulation of ST while maintaining at the same time a full microscopic world-sheet approach. The construction of SFT's is such that ST world-sheet amplitudes are reproduced when these are well-defined. But SFT gives a more general construction of amplitudes which is well-defined even when the standard world-sheet approach gives rise to divergences.  In this very general framework all the elementary string interactions are defined so as to provide a solution to the quantum  Batalin-Vilkovisky  (BV) master equation, furnishing a perturbative microscopic definition of  the target space path integral of ST. This construction is explicitly realized in terms of (quantum) homotopy-algebras for both bosonic strings and superstrings, including Type II, Type I and heterotic.

The construction offered by SFT allows to define the 1PI effective action of string theory and thus to give a definition of string perturbation theory where it is possible to discuss  quantum effects such as vacuum shifts due to tadpoles and mass renormalization.
The explicit knowledge of microscopic UV SFT's  allows to construct the low-energy  ST effective action as the Wilsonian action, by integrating out the massive string states from the SFT path integral. This top-down construction is safe from infrared divergences and recently turned out very useful for obtaining un-ambiguous results on non-perturbative contributions, such as D-instanton corrections to perturbative amplitudes and effective superpotentials.

SFT's (especially open string field theories, OSFT's)  allow to approach background independence in ST by recasting the plethora of different ST backgrounds in the form of classical solutions to the SFT equation of motion. This program has been fully realized in critical  bosonic OSFT, where any D-brane system can be expliclty written as a classical solution of the OSFT on any other D-brane system. 

String Field Theory aims at furnishing a complete formulation of String Theory which can possibly go beyond the perturbative world-sheet description. The subject is rather vast and ramified, so that it is not possible, in a single article, to mention all of the interesting directions that have been pursued in the past fifty years or so. This  article is the result of a choice made by the author to privilege the most recent developments while, at the same time, to still  furnish some important results from the past which are at the basis of today's constructions and approaches. 

The article is articulated in three parts. The first part  (section \ref{chap:2}) deals with the construction of string field theories as a tool to provide an alternative definition of amplitudes, fitting the rules of standard quantum field theory. The second part (section \ref{chap:3}) explains some of the recent results mainly due to A. Sen about the possibility of taming infrared divergences in ST using SFT. Finally the third part (section \ref{chap:4}) deals with the construction of classical solutions in (open) SFT and the (partial) realization of background independence, possibly preparing the ground towards future non-perturbative physics.

This article should be useful for people working in the broad area of String Theory, to get an idea of what String Field Theory is,  together with some of the main directions of current research. The interested reader can sharpen her understanding on the given set of references.
Recent pedagogical reviews  include\cite{deLacroix:2017lif, Erler:2019loq, Erler:2019vhl} \footnote{Other less recent but useful reviews (with a focus on open string field theory) are \cite{Thorn:1988hm, Taylor:2003gn, Sen:2004nf, Okawa:2012ica} } , together with the recent books  \cite{Erbin:2021smf} and \cite{Doubek:2020rbg}.

Throughout this article some familiarity with world-sheet string theory will be assumed, as discussed in classical ST textbooks, for example \cite{Polchinski:1998rq, Polchinski:1998rr}.

\section{Construction of string field theories}\label{chap:2}
The construction of a string field theory (SFT) depends on the particular string theory (ST) which is under consideration. In this contribution  only covariant approaches will be considered (in contrast, for example, to the light-cone approach which started with \cite{Kaku:1974zz, Kaku:1974xu}\footnote{See \cite{Erler:2020beb} for recent work on relating the light-cone theories to the covariant theories.}. All SFT's  share  the important  property that they reproduce, through their  Feynman diagrams, the standard on-shell amplitudes as defined by the Polyakov path integral. The crucial advantage is that while the Polyakov path integral is often divergent in the regions of moduli space where the world-sheet degenerates or, in case of the superstring, where there are spurious singularities \cite{Verlinde:1987sd}, the SFT approach sees  divergences associated to degeneration as space-time infrared divergences and treats them as in ordinary QFT. Moreover it also avoids spurious singularities, by appropriately improving the integration over moduli space thanks to the so-called {\it vertical integration} \cite{Sen:2015hia}.  Therefore SFT provides a complete framework for String Perturbation Theory which is manifestly free of divergences.

A  string field theory action is typically expressed as a (quantum) master action in the sense of  Batalin-Vilkovisky (BV) quantization \cite{Batalin:1981jr, Batalin:1983ggl}\footnote{A useful  pedagogical introduction to BV quantization for SFT purposes  is contained in the classical review by Thorn \cite{Thorn:1988hm} and, more recently, in \cite{deLacroix:2017lif}. See \cite{Henneaux:1992ig} for a complete account of this vast and general subject.}. This action is a functional of the string field $\Phi$, which is a generic state in the Hilbert space of the CFT describing the string background
\begin{align}
\Phi=\sum_s \phi_s\,\ket s+\sum_{ s} \phi^*_s\,\ket {\hat s}.\label{BVSF}
\end{align}
Here $\{\ket s\}$ is a (possibly restricted) basis of half of the full matter-ghost CFT Hilbert space and the coefficients $\{\phi_s\}$ are the  dynamical degrees of freedom collectively representing space-time (BV) fields.  $\{\ket {\hat s}\}$  is a basis of the complementary part of the Hilbert space, such that 
\begin{align}
\langle\!\langle\,\ket{s}\,,\,\ket{\hat r}\,\rangle\!\rangle=\delta_{rs},
\end{align}
where $\langle\!\langle\,\cdot\,,\,\cdot\,\rangle\!\rangle$ is an appropriate inner product  and
 the coefficients $\phi_s^*$ are the (BV) anti-fields. Fields and antifields are graded with respect to the BV degree and a BV anti-field $\phi^*_s$ has a degree which is opposite to the one of corresponding field $\phi_s$. Together with the degree of the basis vectors $\{\ket s ,\ket{\hat s}\}$ given naturally by the world-sheet ghost number, this makes the total BV string field $\Phi$ homogeneously degree even.

 The corresponding BV master action (considering for illustrative purposes closed string field theory) is written as 
\begin{align}
S(\phi_s, \phi_s^*)=S[\Phi]=-\frac1{g_s^2}\left(\frac12\langle\!\langle\Phi,Q\Phi\rangle\!\rangle+\sum_{g=0}^\infty g_s^{2g}\, {\cal V}^{(g)}(\Phi)\right),\label{action}
\end{align}
 where $Q$ is the world-sheet BRST charge. The terms ${\cal V}^{(g)}$ represent the interacting part of the action which is a sum of various contributions associated to genus $g$ Riemann surfaces
 \begin{align}
{\cal V}^{(g)}(\Phi)=\sum_{k=1}^\infty {\cal V}_k^{(g)}(\,\underbrace{\Phi,\cdots,\Phi\,}_{k\text{ times}}\,).\label{Vk}
\end{align}
Notice, first of all, that the action contains a tadpole term $\sum_g {\cal V}_{k=1}^{(g)}$. In standard constructions one starts with an exact CFT background where the sphere ($g=0$) one-point functions vanish, so that  ${\cal V}_{k=1}^{(g=0)}=0$. However, as in any QFT, quantum corrections ($g>0$) will typically induce tadpoles and therefore such terms have to be included in the quantum BV master action. A more detailed discussion on tadpoles and their physical relevance will be carried out in section \ref{chap:3}.

The action \eqref{action} constructed from $(Q,{\cal V})$ must be consistent as a quantum field theory in target space, in the sense that its (BV) path integral must be independent of the choice of gauge fixing. This is stated by the quantum BV master equation
\begin{align}
 \frac12\left\{S[\Phi],S[\Phi]\right\}_{\rm BV}+\Delta_{\rm BV} S[\Phi]=0,
\end{align}
where the BV anti-bracket $\{\cdot,\cdot\}_{\rm BV}$ and the symplectic laplacian $\Delta_{\rm BV}$ are defined as
\begin{align}
\{S,S\}_{\rm BV}&=\sum_s \,S\left(\frac{\overleftarrow\partial}{\partial \phi_s}\frac{\overrightarrow\partial}{\partial\phi^*_s}-\frac{\overleftarrow\partial}{\partial \phi^*_s}\frac{\overrightarrow\partial}{\partial\phi_s}\right)S,\\
\Delta_{\rm BV} S&=\sum_s (-1)^{d(\phi_s)}\frac{\overrightarrow\partial}{\partial\phi_s} \frac{\overrightarrow\partial}{\partial\phi^*_s}S.
\end{align}
The validity of the quantum BV master equation turns out to be associated to the correct covering of the world-sheet moduli space. This in turn crucially depends on the definition of the interaction vertices ${\cal V}_k^{(g)}$,
which are closely related to off-shell amplitudes. Similarly to a standard on-shell amplitude, a $k$-point, genus $g$ off-shell amplitude  is an integral over the  moduli space of a $k$-punctured genus $g$ Riemann surface. The integrand is a correlator of generic (possibly restricted) CFT states inserted at the punctures, together with appropriate anti-ghost $b$ insertions providing the correct measure in moduli space. However, since the inserted states are not in general $(0,0)$ conformal primaries (as it is the case for on-shell amplitudes), they have to be inserted via local coordinate maps, mapping the canonical disk in the complex plane to non-overlapping  patches on the Riemann surface, having the puncture in their interior. This construction clearly depends on the choice of local coordinate maps, which is interpreted in SFT as a freedom of doing field redefinitions. When the external states are physical (BRST closed), the off-shell amplitude reduces to the (unique) on-shell amplitude and BRST exact states decouple.\footnote{See for example \cite{Erler:2019loq} or \cite{deLacroix:2017lif} for a more detailed pedagogical introduction to off-shell amplitudes in string theory.} The interaction vertices ${\cal V}_k^{(g)}$ of SFT are off-shell amplitudes, which are however (and crucially) cut-off at the boundary of moduli space, corresponding to Riemann surface degeneration. The dangerous regions which typically give divergences in the Polyakov path integral (including the collisions of vertex operators) are excluded in the definition of the ${\cal V}_k^{(g)}$, which are then finite for any insertion of the off-shell states.

The full amplitudes are finally constructed from Feynman rules, effectively  gluing the interaction vertices ${\cal V}_k^{(g)}$ with the propagators obtained by the kinetic term. In such a way, after collecting all of the relevant diagrams for a given amplitude, the remaining moduli space integration towards Riemann surface degeneration is performed by the Schwinger parameter $t$ of the propagator 
\begin{align}
\frac{b_0+\bar b_0}{L_0+\bar L_0}=(b_0+\bar b_0)\int_0^\infty dt\, e^{-t(L_0+\bar L_0)},
\end{align}
which precisely creates the `long tubes' associated to the degenerating regions.

Crucially, the vertices ${\cal V}_k^{(g)}$ have to be carefully constructed in such a way that the amplitudes obtained by summing up the Feynman diagrams correctly integrate  over the full moduli space, without overcounting or missing any region. This is encoded  \cite{Sen:1994kx} in the fundamental equation
\begin{align}
\partial {\cal V}+\frac12\{{\cal V},{\cal V}\}+\Delta {\cal V}=0,\label{geom-bv}
\end{align}
where ${\cal V}$ is a formal sum packaging all the vertices at given genus and all possible number of punctures
\begin{align}
&{\cal V}=\sum_{g}g_s^{2g}\,{\cal V}^{(g)},\\
&{\cal V}^{(g)}=\sum_k\,{\cal V}^{(g)}_k.
\end{align}
A vertex ${\cal V}^{(g)}$ integrates in a region of moduli space of a genus $g$ punctured surface and the meaning of \eqref{geom-bv} is that the boundary of this region (computed by $\partial$) should match the  gluing  of two sub-vertices (captured by the term $\{{\cal V},{\cal V}\}$), if the boundary is related to a separating degeneration, or the gluing of two punctures from the same surface (represented by the term $\Delta {\cal V}$) if the boundary is related to a non-separating degeneration. This is indeed required by the decoupling of BRST exact states from the full amplitude constructed from the vertices ${\cal V}$'s and the propagators.
The vertices ${\cal V}$ are not uniquely defined by \eqref{geom-bv}, but it can be shown that the actions constructed from ${\cal V}$ and ${\cal V}'$ both satisfying \eqref{geom-bv} are related by field redefinition and are thus equivalent. 
Constructing explicit examples of these vertices is highly non-trivial and continues to be an important constant endeavour  in the SFT research \cite{Saadi:1989tb, Zwiebach:1992ie, Zwiebach:1997fe, Cho:2019anu, Firat:2021ukc, Costello:2019fuh,  Erbin:2022rgx, Firat:2023glo, Firat:2023suh  }. 

  It is  one of the core-properties of the above geometrical construction \cite{Zwiebach:1992ie, Sen:1994kx} that the quantum BV master equation is a consequence of the `geometrical' BV master equation \eqref{geom-bv}
\begin{align}
\partial {\cal V}+\frac12\{{\cal V},{\cal V}\}+\Delta {\cal V}=0\quad\to\quad \frac12\left\{S[\Phi],S[\Phi]\right\}_{\rm BV}+\Delta_{\rm BV} S[\Phi]=0.
\end{align}
In the following some of the details of this construction will be presented in specific examples.
\subsection{Bosonic String Field Theory}
 
In bosonic string theory the world-sheet CFT consists of a {\it matter} $c=26$ CFT and the universal ghost $bc$ system with $c=-26$. In the critical realization, the $c=26$ theory is given by $26$  bosons $X^\mu(z,\bar z)$ possibly interacting  in a sigma-model, maintaining conformal invariance. In the non-critical realization the matter CFT is realized by a generic $c$ CFT, compensated by a  $26-c$
Liouville CFT.  For $c\leq 1$ the theory is free from tachyonic instabilities and it makes sense to study it quantum mechanically. Indeed it turns out that in this case the full string theory is equivalently described by a matrix model, for review see \cite{Klebanov:1991qa, DiFrancesco:1993cyw}.

\subsubsection{Open Bosonic SFT}
Open bosonic SFT  is the simplest realization of a {\it classical} SFT. To define it, it is first of all necessary to focus on a possible open string spectrum of a consistent closed string background described by a reference CFT$_0$. This means to choose a reference D-brane system described by a reference boundary CFT (BCFT), BCFT$_0$, with a corresponding Hilbert space ${\cal H}_{\rm o}$ of boundary fields. 
The open BV string field $\Psi$ is a generic state in ${\cal H}_{\rm o}$. On this space there exists a non-degenerate inner product (the standard (B)CFT inner product) 
\begin{align}
\langle\,\cdot\,,\,\cdot\,\rangle:\,{\cal H}_{\rm o}\otimes{\cal H}_{\rm o}\to \mathbb{C},
\end{align}
which computes the boundary two-point functions and which can be used to define a free action functional 
\begin{align}
S_{\rm free}[\Psi]=-\frac1{2g_s}\langle\Psi,Q_{B}\Psi\rangle,
\end{align}
where $Q_B$ is the nilpotent world-sheet BRST charge.\footnote{The possibility of using the world-sheet BRST symmetry to write down a space-time kinetic term has been originally explored in \cite{Siegel:1984ogw, Siegel:1985tw, Banks:1985ff, Itoh:1985bb}. The final gauge-unfixed expression $\langle\Psi|Q|\Psi\rangle$ was reported in \cite{Neveu:1985sh } and \cite{Witten:1985cc}.} The equations of motion of this free action
\begin{align}
Q_B\Psi=0,
\end{align}
 correctly reproduce the open string spectrum, together with the gauge invariance
 \begin{align}
\Psi\sim\Psi+Q_B\Lambda.
\end{align}
The next step is to introduce {\it interactions}
 \begin{align}
S[\Psi]=-\frac1{g_s}\left(\frac12\langle\Psi,Q_{B}\Psi\rangle+\sum_{n=2}^\infty\frac1{n+1}\langle\Psi,m_n(\,\underbrace{\Psi,\cdots\!,\Psi}_{n\,\text{times}})\rangle\right),
\end{align}
where the products have been defined
\begin{align}
m_n:\,{\cal H}_{\rm o}^{\otimes n}\to{\cal H}_{\rm o},\label{m_n}
\end{align}
which, together with the inner product, realize the couplings ${\cal V}_{n+1}$ defined in \eqref{Vk}, with the understanding that in this case the corresponding Riemann surface is a  disk.
The consistency requirement is that the Feynman diagrams of this theory should correctly cover the world-sheet moduli space of boundary-punctured disks, which are the Riemann surfaces associated to classical open strings. As anticipated, this can be  stated by asking that the (classical) Batalin-Vilkovisky (BV) master equation is obeyed
\begin{align}
\{S[\Psi],S[\Psi]\}_{\rm BV}=0.
\end{align}
The classical BV master equation is in turn satisfied if the products $m_n$ obey the so-called $A_\infty$ relations \cite{Gaberdiel:1997ia}
\begin{align}
\sum_{k=1}^{n-1}m_k m_{n-k}=0,
\end{align}
where $m_1\coloneqq Q_B$ and 
\begin{align}
m_km_p:&\,\,{\cal H}_{\rm o}^{\otimes k+p-1}\to {\cal H}_{\rm o}\\
m_km_p(\Psi_1,\cdots,\Psi_{k+p-1})\coloneqq& m_k(m_p(\Psi_1,\cdots,\Psi_p),\Psi_{p+1},\cdots\Psi_{k+p-1})\0\\
+(-1)^{d_1}&m_k(\Psi_1,m_p(\Psi_2\cdots,\Psi_{p+1}),\Psi_{p+2}\cdots\Psi_{k+p-1})\0\\
&\vdots\0\\
+(-1)^{d_1+\cdots d_{k-1}}&m_k(\Psi_1,\cdots\Psi_{k-1},m_p(\Psi_k\cdots,\Psi_{k+p-1})),
\end{align}
where $d_i$ is the degree of $\Psi_i$ which is defined as the opposite of its grassmannality (which is in turn given by the sum of the world-sheet ghost number and the BV grading).

It turns out that there exists a very peculiar choice  of the multistring products $m_n$ where the full moduli space is covered with only the propagator $b_0/L_0$ and an associative two-string product $m_2$. This is the celebrated Witten OSFT \cite{Witten:1985cc}, which is the only covariant string field theory which is purely cubic
\begin{align}
S_{\rm Witten}[\Psi]=-\frac1{g_s}\left(\frac12\langle\Psi,Q_{B}\Psi\rangle+\frac13 \langle\Psi,m_2(\Psi,\Psi)\rangle\right).
\end{align}
The simplicity of this theory, as compared to the other string field theories, allowed several advances in the non-perturbative understanding of the space of open string vacua, which will be partly discussed in section \ref{chap:4}.

As far as quantum open strings are concerned, it is possible to formally extend the $A_\infty$ algebra to a quantum $A_\infty$ algebra, so as to provide a solution to the quantum BV master equation for pure open stings. However it is not clear how this  relates to  the closed string poles which are generated in the two open string non-planar one-loop diagram \cite{Shapiro:1987ac, Freedman:1987fr}, which require closed strings as external states to preserve unitarity.  In the case of Witten theory this can be formally achieved by deforming the initial action with gauge invariant tadpoles  which couple on-shell closed strings $V^k$ to a dynamical open string
\begin{align}
S^{(\mu)}_{\rm Witten}[\Psi]=-\frac1{g_s}\left(\frac12\langle\Psi,Q_{B}\Psi\rangle+\frac13 \langle\Psi,m_2(\Psi,\Psi)\rangle\right)+\sum_k\mu_k \langle V^k(i,-i), \tilde \Psi\rangle,
\end{align}
where $\tilde\Psi$ is an appropriate conformal transformation of the original dynamical open string field $\Psi$ which can be obtained from the non planar one-loop amplitude in the region of closed string degeneration \cite{Shapiro:1987ac} and where $V^k(i,-i)$ represents the insertion of the closed string vertex operator $V^k(z,\bar z)$ at the open-string midpoint $z=i$. Using these operators it is possible to formally reproduce all open-closed amplitudes covering the full moduli space with just open string propagators \cite{Zwiebach:1992bw}, also in the regions of closed string degeneration. Notice that the addition of these gauge invariant terms to the OSFT action is analogous to the deformation of a gauge theory by gauge invariant operators. These gauge invariant operators (often called `Ellwood Invariants') have been also very useful to explicitly construct the boundary state associated to OSFT classical solutions \cite{Ellwood:2008jh, Kudrna:2012re} and they have been recently used to characterize D-branes deformation to first order in a continuous  change of the closed string background \cite{Maccaferri:2021ksp, Maccaferri:2021lau, Maccaferri:2021ulf}.  They are expected to play a role in a possible quantum structure of Witten OSFT. \footnote{Interesting ideas on the possible quantum consistency of OSFT alone  have been formulated in the form of the  `open string completeness conjecture' (see \cite{Sen:2004nf}) and more recently in \cite{Okawa:2020llq}. In another direction however, the inclusion of propagating off-shell closed strings resulted  necessary  to construct (and compute) well-defined D-branes' observables \cite{Maccaferri:2021ksp, Maccaferri:2022yzy}.  }
In a generic string background a quantum formulation of open strings is nevertheless always possible in the complete framework of open-closed SFT \cite{Zwiebach:1990qj, Zwiebach:1997fe}. 

\subsubsection{Closed Bosonic SFT}
The next theory to analyze in order of complexity is the closed bosonic SFT, formulated in \cite{Zwiebach:1992ie}. The definition starts with a choice of exact bosonic string background consisting of a matter CFT$_{\rm m}$ with $c=26$, together with the $bc$ CFT$_{\rm gh}$. This time the Hilbert space ${\cal H}_{\rm c}$ is made up of bulk fields $\Phi$ of CFT$_{\rm m}\otimes$CFT$_{\rm gh}$, importantly subject to the level matching constraints
\begin{align}
(b_0-\bar b_0)\Phi&\coloneqq b_0^-\Phi=0\\
(L_0-\bar L_0)\Phi&\coloneqq L_0^-\Phi=0.
\end{align}
These constraints allow to write a consistent kinetic term as
\begin{align}
S_{\rm free}[\Phi]=-\frac1{2g_s^2}\,\langle\Phi,c_0^-Q_B\Phi\rangle,
\end{align}
where $c_0^-=1/2(c_0-\bar c_0)$ and this time $\langle\,\cdot\,,\,\cdot\,\rangle$ is the BPZ inner product on the Riemann sphere.
This gives the correct physical state condition and the linearized off-shell gauge invariance
\begin{align}
Q_B\Phi&\,=0\\
\Phi\sim&\,\Phi+Q_B\Lambda,\\
\end{align}
where $b_0^-\Lambda=L_0^-\Lambda=0$.
The extension to an interacting classical theory is analogous to the open string case and reads 
\begin{align}
S_{\rm class}[\Phi]=-\frac1{g_s^2}\left(\frac12\langle\Phi,c_0^-Q_{B}\Phi\rangle+\sum_{k=2}^\infty\frac1{(k+1)!}\langle\Phi,c_0^-l_k(\Phi^{\wedge k})\rangle\right).
\end{align}
The $l_n$'s are multi-string products
\begin{align}
l_n:\,{\cal H}_{\rm c}^{\wedge n}\to{\cal H}_{\rm c},
\end{align}
which are graded-symmetric in their input ($\wedge$ is the symmetrized tensor product) and are cyclic with respect to the inner product $\aver{\cdot\,,c_0^-\,\cdot}$. As for the previous open string case, the products should be constructed in such a way that the Feynman diagrams cover the moduli space of the relevant surfaces, in this case punctured spheres. Again this is formally guaranteed by the classical BV master equation
\begin{align}
\{S_{\rm class}[\Phi],S_{\rm class}[\Phi]\}_{\rm BV}=0.
\end{align}
The classical BV master equation is in turn satisfied if the products $l_n$ obey the so-called $L_\infty$ relations
\begin{align}
\sum_{k=1}^{n-1}l_k l_{n-k}=0,\label{l-inf}
\end{align}
where $l_1\coloneqq Q_B$. In the case of closed strings it is possible to extend the classical construction to a full (perturbative) quantum theory without necessarily introducing open strings. This is done by adding new interaction vertices corresponding to punctured genus $g$ surfaces
\begin{align}
l_n\to l_n^{(g)}, \quad g=0,1,\cdots\,.
\end{align}
The quantum corrected action has the structure
\begin{align}
S_{\rm quant}[\Phi]=-\sum_{g=0}^\infty\,g_s^{-2+2g}\sum_{k=0}^{\infty}\frac1{(k+1)!}\langle\Phi,c_0^-l_k^{(g)}(\Phi^{\wedge k})\rangle.
\end{align}
 This time the requirement of the correct covering of moduli space of punctured genus $g$ Riemann surfaces requires the action to obey the quantum BV master equation
 \begin{align}
 \frac12\{S_{\rm quant}[\Phi],S_{\rm quant}[\Phi]\}_{\rm BV}+\Delta_{\rm BV} S_{\rm quant}[\Phi]=0.
 \end{align}
This equation is satisfied provided  the products $l_k^{(g)}$ obey appropriate generalizations of \eqref{l-inf} which are called  quantum $L_\infty$ relations, \cite{Zwiebach:1992ie, Markl:1997bj}, whose role is to make sure that the amplitudes computed from the Feynman rules of the quantum BV action are consistent.

\subsubsection{Open-Closed Bosonic SFT}
When D-branes and open strings are involved in a closed string process,  it is possible to consider a complete open-closed SFT \cite{Zwiebach:1990qj, Zwiebach:1997fe}. This theory can be considered as a non-perturbative completion of closed string field theory in the sense that D-branes contribution to closed string amplitudes should be contained thanks to the dynamical open string degrees of freedom.  The free theory is the sum of the open and closed free theories
\begin{align}
S^{oc}_{\rm free}[\Phi,\Psi]=-\frac1{2g_s^2}\,\langle\Phi,c_0^-Q_B\Phi\rangle-\frac1{2g_s}\langle\Psi,Q_{B}\Psi\rangle.
\end{align}
Differently from the purely open or purely closed SFT's, this time the interacting theory does not exist at a purely classical level but it is directly defined at the level of the quantum BV master action
\begin{align}
 S^{oc}_{\rm quant}[\Phi,\Psi]=\sum_{g,b=0}^{\infty}g_s^{2g+b-2}\sum_{k=0}^{\infty}\sum_{\{l_{1},...,l_{b}\}=0}^{\infty}\mathcal{A}^{g,b}_{k;\{l_{1},...,l_{b}\}}\left(\Phi^{\wedge k}\otimes'\Psi^{\odot l_{1}}\wedge'...\wedge' \Psi^{\odot l_{b}}\right).\label{OC-action}   
\end{align}
The interaction vertices $\mathcal{A}^{g,b}_{k;\{l_{1},...,l_{b}\}}$ are off-shell amplitudes on Riemann surfaces of genus $g$ and $b$ boundaries, where $k$ off-shell closed strings $(\Phi)$ are inserted in the bulk in a permutational invariant way. In addition, for every boundary $i=(1,\cdots,b)$, $l_i$ open strings $(\Psi)$  are cyclically inserted ($\odot$ is the cyclic tensor product), with complete permutational symmetry between the boundaries (represented by the symmetrized tensor products $\wedge'$). These off-shell amplitudes are integrated over moduli space with a cutoff towards open and closed string degeneration in such a way that the full S-matrix will be composed by these amplitudes, together with the Feynman diagrams obtained by connecting sub-amplitudes with open and closed string propagators. The consistency of this construction (for example the decoupling of BRST exact states in on-shell amplitudes) is encoded in the quantum BV master equation
 \begin{align}
\{S^{oc}_{\rm quant},S^{oc}_{\rm quant}\}_{\rm c}+ \{S^{oc}_{\rm quant},S^{oc}_{\rm quant}\}_{\rm o}+2\Delta_{\rm c}S^{oc}_{\rm quant}+2\Delta_{\rm o}S^{oc}_{\rm quant}=0.\label{quantum-BV-closed}
 \end{align}
Also in this case the fundamental BV structures can be realized via multistring open-closed products obeying an open-closed quantum homotopy algebra \cite{Kajiura:2004xu, Munster:2011ij, Maccaferri:2022yzy}. Recently it has been possible to recast this open-closed quantum homotopy structure into an equivalent nilpotent structure defined on an appropriate open-closed tensor algebra \cite{Maccaferri:2023gcg}, with the advantage of simplifying several algebraic manipulations.

\subsection{Overview of Superstring Field Theory}
The most interesting applications of string theory are in the framework of the superstring.  The microscopic structure of superstring theory is richer and more complicated than the bosonic string, although the general idea that amplitudes are associated to sum over two dimensional surfaces is still foundational. However new structures (or new conformal field theories) are added  to the  Riemann surfaces which are used in the bosonic string.
Superstring theory has mainly two distinct covariant world-sheet formulations. 

The first and the oldest is the Ramond-Neveu-Schwarz (RNS), which arises as the gauge fixing of $N=(1,1)$ world-sheet supergravity, coupled to world-sheet matter \cite{Polchinski:1998rr}. This formulation has explicit world-sheet supersymmetry but space-time supersymmetry is less explicit and it is fully realized only modulo world-sheet supersymmetry (technically modulo {\it picture changing}) \cite{Friedan:1985ge}. 

The second formulation is the Pure-Spinor (PS) formulation \cite{Berkovits:2000fe} which has the advantage of being explicitly space-time supersymmetric in ten dimensions. The origins of the PS theory are however less understood than  RNS, although connections between the two approaches have been established \cite{Berkovits:2005bt, Berkovits:2005ng,     Berkovits:2013eqa} . The second-quantized version of the PS open string has been addressed in \cite{Berkovits:2005bt, Kroyter:2012xk} but subtleties related to the conical singularity and the non-compactness of the pure-spinor space have remained to be satisfactorily understood \cite{Aisaka:2008vw, Aisaka:2009yp, Bedoya:2009np}. 

On the other hand,  much progress has been achieved in recent years in the context of the RNS superstring, also triggered by Witten's revisitation of this old subject \cite{Witten:2012bh, Witten:2013cia}. Even in the context of RNS there are still different related approaches that are considered both in first and in second quantization. To appreciate these different approaches let's consider the construction of superstring amplitudes. Formally, they should be understood as integral forms to be integrated over the supermoduli space of super-Riemann surfaces with NS or R (open or closed) punctures. The first and most geometric  approach to the RNS string is precisely at the level of giving a concrete meaning to the integration theory on the supermoduli space \cite{DHoker:1988pdl, Witten:2012bh}. At the second-quantized level, the geometrical super-moduli picture has been preliminarily explored in \cite{Ohmori:2017wtx} in the context of open superstring field theory but there has not been much  progress since then.

 The second possible approach consists in first integrating over the fermionic directions of the super-moduli space. This has the effect of inserting extra operators (the picture-changing operators, PCO's) in the remaining integral over the bosonic moduli,  \cite{Verlinde:1987sd, Witten:2012bh}. However it turns out that the fermionic integration cannot be done globally over the bosonic moduli space \cite{Donagi:2013dua} but, in general, it has to be done patch by patch. This is also associated to the appearence of  `spurious singularities' in the superghost correlation functions \cite{Verlinde:1987sd}. This results in a distribution of PCO's which varies discontinuously by changing patch in the bosonic moduli space. To correct for these discrepancies (which would result in breakdown of BRST decoupling in amplitudes)  compensating terms should be added,  which have the effect of moving the PCO's without changing the bosonic moduli. This is an example of  {\it vertical integration} \cite{Sen:2015hia}. This ``partially integrated'' picture is the one that has been mostly studied in the context of superstring field theory.  
 
 At the level of the world-sheet path-integral the insertion of the PCO's has the effect of giving the needed insertions of the delta functions $\delta(\gamma)$ and $\delta(\beta)$ to integrate over the bosonic non-compact zero-modes of  $(\beta, \gamma)$  which, if not localized by the delta functions,  would give divergences.
 This brings  to the third approach to RNS which is more technical in nature (and which is  the least understood from a second quantized perspective) where one bosonizes the $\beta \gamma$ system as $\gamma=e^\phi\eta$ and $\beta=\partial\xi e^{-\phi}$. In this way the somewhat unusual world-sheet fields $(\delta(\gamma), \delta(\beta))$ are replaced by the more standard conformal fields $(e^{-\phi},e^{\phi})$. As it is well-known however,  this is a subtle replacement because the $\eta\xi\phi$-path integral is also done over the $\xi$ zero mode, which does not enter into the definition of $\beta$ and $\gamma$. This gives rise to a subtle doubling of world-sheet degrees of freedom and the notion of {\it Large Hilbert Space} (LHS, where the $\xi$ zero-mode is present) versus {\it Small Hilbert Space} (SHS, which is truly isomorphic to the $\beta\gamma$ Hilbert space) \cite{Friedan:1985ge}. As of today it is possible to construct classical gauge invariant superstring field theories in the LHS but their BV structure remains rather elusive \cite{Kroyter:2012ni, Berkovits:2012np}. 
 
One of the major difficulties in writing down a superstring field theory in the RNS approach has  been the construction of a kinetic term for Ramond fields. Recently this has been resolved in two  independent ways: either by restricting the space where off-shell Ramond fields are defined \cite{Kunitomo:2015usa} or by adding extra non-interacting Ramond fields \cite{Sen:2015hha, Sen:2015uaa} which are needed to construct the quadratic term of the action but which decouple from interactions. 

It follows a concise summary and a guide through the literature for the various available RNS covariant superstring field theories

Classical open superstring field theory has been formulated in a Wess-Zumino-Witten (WZW) like form in the LHS in the NS sector in \cite{Berkovits:1995ab}. The same algebraic structure was also shown to accomodate a full super-Poincar\'e invariant action in 4 dimensions using the so-called `hybrid' world-sheet formulation.\footnote{The very flexible WZW structure has also been useful to construct a LHS formulation of Heterotic String Field Theory in the NS sector \cite{Berkovits:2004xh}.} This LHS RNS theory has been later upgraded to contain the Ramond sector in \cite{Kunitomo:2015usa}. This resulted in the first classical RNS open superstring field theory containing both the NS and the R sector. Via a partial gauge fixing and field redefinition this construction has been recast in the SHS with an explicit cyclic $A_\infty$ structure \cite{Erler:2016ybs}.   An important part of the construction in the SHS has been how to distribute the PCO's in the fundamental vertices in such a way that there are no dangerous collisions and, at the same time, the $A_\infty$ relations are satisfied. This has been a major contribution of   \cite{Erler:2013xta}. 


As far as closed superstring field theory is concerned, complete  ({\it i.e.} with a proper treatment of the R sector) classical RNS actions  are available for the heterotic string \cite{Kunitomo:2019glq} and Type II superstrings \cite{Kunitomo:2019kwk}, where the necessary distribution of PCO's has been constructed using homotopy algebra methods analogous to \cite{Erler:2013xta}. The algebraic structure of a quantum closed superstring theory has been elucidated in \cite{Sen:2015uaa}. However a  complete and explicit construction of the needed PCO distributions at loop level, fully dealing with spurious singularities, has not  been written down yet.

Similar considerations apply to open-closed super SFT whose quantum algebraic structure  has been described in \cite{FarooghMoosavian:2019yke} and explicitly realized  by appropriately  distributing the PCO's, at the level of spheres and disks, in \cite{Kunitomo:2022qqp, jakub-talk}.
\section{Infrared-safe String Perturbation Theory}\label{chap:3}
The amplitudes obtained by SFT Feynman diagrams  reproduce by construction the usual amplitudes as defined by the Polyakov path integral, {\it when such an integral is well defined}. String theory is very soft at high energy and this results in the absence of UV divergences. However ST contains massless particles and therefore it is potentially afflicted with infrared divergences just like any other theory with massless particles. These singularities  show up as divergences in the moduli space integral associated to Riemann surfaces close to degeneration. In the SFT approach this region is always associated to a Feynman diagram with an explicit closed or open propagator and this has the consequence that the associated divergences   can be dealt with by a stringy generalization of the ``$i\epsilon$'' prescription \cite{Witten:2013pra}. In particular it can be explictly seen that these divergences are always associated to the propagation of  tachyonic or massless degrees of freedom through the degenerating neck (closed strings) or strip (open strings) in the corresponding surfaces.
\subsection{1PI effective action and mass renormalization}

One of the problems which is typically associated with infrared divergences is that of loop tadpoles which in a generic quantum field theory destabilize the classical perturbative vacuum. In string theory the classical perturbative vacuum is given by a choice of world-sheet CFT. Depending on the CFT it can be that the one-loop tadpole amplitude is not vanishing and thus the vacuum has to be shifted in a way which depends  on the string coupling constant. In a SFT setting this means
\begin{align}
\Phi_{\rm v}=0\quad\to\quad \Phi_{\rm v}=\Phi_{\rm v}(g_s).
\end{align}
Notice that while $\Phi_{\rm v}=0$ represents a two-dimensional CFT, this is not true anymore for the shifted vacuum represented by $\Phi_{\rm v}(g_s)$. The vacuum shift solution $ \Phi_{\rm v}(g_s)$ can be obtained by extremizing   the 1PI-effective action \cite{Sen:2014dqa, Sen:2015hha}.
 The couplings defining the 1PI-effective action are similar to the ones defining the BV action, i.e. they are off-shell amplitudes with a cut-off towards  degenerations. However  this time the cutoff is only present for {\it separating} degenerations. Alternatively, it can be constructed, as in standard QFT,  by computing 1PI off-shell amplitudes with the Feynman rules inherited from the quantum BV master action.  Considering for simplicity a closed string field theory, these couplings will be naturally expanded in the genus as
\begin{align}
l_k^{(1PI)}=\sum_{g=0}^\infty \, {g_s}^{2g} l_k^{(1PI,g)}
\end{align}
and the 1PI-effective action will have the form
\begin{align}
S^{(1PI)}(\Phi)=\frac1{g_s^2}\sum_{k=0}^\infty\,\frac1{(k+1)!}\langle\Phi,c_0^-l_k^{(1PI)}(\Phi,\cdots\!,\Phi)\rangle,
\end{align}
which can be shown to satisfy the classical master equation
\begin{align}
\{S^{(1PI)}(\Phi)),S^{(1PI)}(\Phi)\}_{\rm BV}=0.
\end{align}
This action inherits from the quantum BV master action the presence of a tadpole which destabilizes the classical perturbative vacuum $\Phi=0$. This tadpole appears as  a zero-product $l_0^{(1PI)}$ and is given by a genus expansion, starting at 1-loop
\begin{align}
l_0^{(1PI)}=\sum_{g=1}^\infty {g_s}^{2g}\,l_0^{(1PI,g)}.
\end{align}
The presence of the tadpole does not spoil the fact that the 1PI effective action obeys the classical BV master equation, but this time the homotopy relations obeyed by the multi-string products $l_k^{(1PI)}$ define a so-called {\it weak} $L_\infty$ algebra
\begin{align}
\sum_{k=0}^n l_k^{(1PI)}l_{n-k}^{(1PI)}=0.
\end{align}
Notice in particular that the quadratic term (the kinetic operator) is no more nilpotent
\begin{align}
\left(l_1^{(1PI)}\right)^2=-l_2^{(1PI)}l_0^{(1PI)}\neq0
\end{align}
and therefore it seems that the notion of physical fluctuations is lost.
These problems (the `classical' tadpole and the breakdown of BRST symmetry) are solved simultaneously  by shifting the perturbative vacuum in such a way as to cancel the tadpole, in a way that is analogous to standard QFT. This requires to solve the {\it vacuum-shift} equation
\begin{align}
\sum_{k=1}^\infty \frac1{k!}l_k^{(1PI)}(\Phi^{\wedge k})=-l_0^{(1PI)}.
\end{align}
The solution $\Phi=\Phi_{\rm v}$ can be searched perturbatively in $g_s$ in a genus expansion\footnote{In general it is also possible to consider different power series expansions.}
\begin{align}
\Phi_{\rm v}(g_s)=\sum_{g=1}^{\infty}(g_s)^{2g} \Phi_{g},
\end{align}
solving the recursive relations
\begin{align}
O(g_2^2):\quad Q\Phi_1&=-l_0^{(1PI,1)}\\
O(g_2^4):\quad Q\Phi_2&=-l_0^{(1PI,2)}-l_1^{(1PI,1)}(\Phi_1)-\frac12 l_2^{(1PI,0)}(\Phi_1,\Phi_1)\\
&\vdots\,.
\end{align}
In general however, it is not guaranteed that these recursive equations have a solution and this is related to the possible presence of massless tadpoles. This is clear by looking at the first order equation which admits a solution only if the one-loop tadpole $l_0^{(1PI,1)}$ is BRST exact. Assuming flat Minkowski space as a background, the one-loop tadpole has obviously zero momentum and therefore the obstruction to solve the vacuum shift to first order is manifested by the presence of zero-momentum cohomology in $l_0^{(1PI,1)}$. Similar obstructions can arise at higher genus by analogous mechanisms. If at some  genus a non trivial element of the zero-momentum cohomology is found this means that the vacuum shift cannot be found perturbatively in $g_s$ and this perturbative scheme breaks down. This is expected to be the case for example for NS-NS tadpoles in certain string compactifications breaking supersymmetry which are expected to present a run-away potential where the new stable vacuum is at infinite distance in field space \cite{Dudas:2004nd}. As of today, there is not the technology to treat this problem neither in SFT, nor in ordinary QFT. On the other hand it is possible to successfully treat the case where, order by order, the obstructions are vanishing  and the vacuum shift will thus appear to be parametrically close to the initial perturbative vacuum $\Phi=0$, so that $\lim_{g_s\to 0}\Phi_{\rm v}(g_s)=0$. This is the case, for example,  of the Fayet-Ilioupulos term generated by compactifying the $SO(32)$ heterotic superstring on a Calabi-Yau 3-fold, which breaks supersymmetry, which is however restored at a nearby dynamically shifted vacuum \cite{Sen:2015uoa}.

Assuming the tadpole has been successfully canceled by the vacuum shift solution $\Phi_{\rm v}$, the theory around the dynamically shifted vacuum will be obtained by writing the original string field $\Phi$ as 
\begin{align}
\Phi=\Phi_{\rm v}+\hat\Phi
\end{align}
Then by expanding the 1PI action around the vacuum shift solution, the tadpole disappears by construction and a new kinetic term arises
\begin{align}
S_2[{\hat\Phi}]=\frac1{2g_s^2}\langle\hat\Phi, l_1(g_s) \hat\Phi\rangle.
\end{align}
The kinetic operator is now nilpotent and its cohomology represents the quantum corrected physical states. In this way one can systematically compute mass renormalizations in string theory, order by order in $g_s$. 

\subsection{Wilsonian effective action}\label{chap:33}

SFT is a QFT for an infinite set of spacetime fields and in several physical applications one is interested only in a set of these degrees of freedom, for example the light fields. In this case a description with the other massive fields integrated out is useful.
The possibility of constructing a low energy effective action from SFT has been explored since \cite{Gross:1986mw, Taylor:2000ek, Berkovits:2003ny}, as an alternative approach to the more traditional methods, for example \cite{Fradkin:1985qd, Fradkin:1985ys, Koerber:2002zb }. More recently Sen has given a concrete construction of the Wilsonian effective action \cite{Sen:2016qap}, which has been followed up by  \cite{Maccaferri:2018vwo, Maccaferri:2019ogq, Erbin:2019spp,  Erbin:2020eyc, Koyama:2020qfb, Masuda:2020tfa} which also clarified the algebraic structure of the effective action in terms of homotopy algebras.

 In order to construct the Wilsonian effective action one starts with a projector $P$, commuting with the BRST charge $Q_B$ which projects on the states in the total off-shell Hilbert  space ${\cal H}$ which should contain the dynamical fields appearing in the effective action
\begin{align}
P: {\cal H}\to&\, P{\cal H}\\
[Q_B,P]&=0.
\end{align}
For example,  to keep the massless states of the closed bosonic string it is possible to take $P$ to be the projector into the kernel of the level operator $\hat L=L_0^+ -\frac{\alpha' k^2}{2}$.
The Wilsonian effective action then has the same form as the quantum BV master action \eqref{quantum-BV-closed} where the dynamical fields are now states in $P{\cal H}$ and  where there are now new products
\begin{align}
l_k^{(g)}\to \tilde l_k^{(g)},
\end{align}
whose structure reflects the fact that the massive fields (the states in the kernel of $P$) have been integrated out
\begin{align}
S_{\rm eff}[\phi]=-\sum_{g=0}^\infty\,g_s^{-2+2g}\sum_{k=0}^{\infty}\frac1{(k+1)!}\langle\phi,\,c_0^-\,\tilde l_k^{(g)}(\phi,\cdots\!,\phi)\rangle\label{Seff}.
\end{align}
The structure of $\tilde l_k^{(g)}$ is that of a full off-shell amplitude computed in the microscopic SFT with a choice of gauge to define the propagator, for example Siegel gauge $b_0^+=0$, \cite{Siegel:1984ogw}, where the propagator is $\frac {b_0^+}{L_0^+}$. Crucially however the states in $P{\cal H}$ are removed from internal lines by an explicit insertion of $\bar P=1\!-\!P$ next to $\frac {b_0^+}{L_0^+}$. The effective couplings of the Wilsonian action are therefore full off-shell amplitudes with the `light' fields $\phi$  as external states and with 
\begin{align}
h\coloneqq \frac {b_0^+}{L_0^+}\bar P
\end{align}
as propagator. 
The action \eqref{Seff} does not loose any information on the high-energy degrees of freedom that have been integrated out and it can be used to compute scattering amplitudes between the $\phi$ fields to any given order in perturbation theory. This is guaranteed by the fact that the effective action \eqref{Seff} still solves the quantum BV master equation with respect to the light fields $\phi$
\begin{align}
\frac12\{S_{\rm eff}(\phi),S_{\rm eff}(\phi)\}_{\rm BV}+\Delta_{\rm BV} S_{\rm eff}(\phi)=0.
\end{align}
\subsubsection{Applications to D-instanton effects}
 A notable example where the approach given by the  Wilsonian effective action has been very fruitful  consists in the computation of D-instanton contributions to closed string amplitudes. These are non-perturbative  $e^{-n/g_s}$ effects which are associated with the virtual creation and annihilation of D-branes (D-instantons) in closed string scattering and, on the world-sheet, they appear as surfaces with boundaries with Dirichlet conditions \cite{Polchinski:1994fq}. These contributions are not automatically generated by the Polyakov path integral, which only gives the perturbative part of a string amplitude, but their presence can be inferred, for example in Type-IIB, by the requirement of S-duality invariance \cite{Green:1997tv}, or in the case of the $c=1$ non-critical string, by comparison with the matrix model.  Precisely in this context \cite{Balthazar:2019rnh} have considered ZZ-instantons contributions to closed string scattering from a genuine world-sheet perspective, using the dual matrix-model description for comparison. The world-sheet computation consists in adding amplitudes with ZZ-boundaries to the standard genus expansion, weighting them by a normalization proportional to the exponential of the D-instanton action $e^{-1/g_s}$. For these peculiar amplitudes, where the boundaries have strict Dirichlet conditions, the integration in moduli space is usually divergent in the region of open string degeneration. This corresponds to regions in moduli space where one or more closed string punctures get close to the boundary or two boundaries get very close to each other. In all of these cases this is conformally equivalent to the formation of one or more very long strips of world-sheet. Divergences arise when tachyonic or massless open string states propagate through these degenerate strips. These divergences are rather common in perturbative string amplitudes and they are generically treated by tuning the external momenta in such a way that the internal lines are not on the mass-shell.  However open strings on D-instantons  have strictly zero momentum and therefore it is not possible to use analytic continuation in external momenta to regulate these degenerations. In the end, by the Fishler-Susskind mechanism \cite{Fischler:1986ci, Fischler:1986tb}, these divergences cancel between  diagrams of different topology. However, since every diagram is independently regulated, the remaining finite parts remain ambiguous in usual string perturbation theory. In \cite{Balthazar:2019rnh} these ambiguities have been fixed by comparing with the matrix model, without a genuine world-sheet justification. In this regard the geometric regularization offered by string field theory \cite{Larocca:2017pbo, Sen:2019jpm}  turned out to be fundamental to fix the ambiguities \cite{Sen:2019qqg}.
 This was achieved by computing the Wilsonian effective action, integrating out completely the massive open string modes on the D-instantons and keeping the massless D-instanton zero-modes as external states, to be integrated at the very end, after having evaluated the open-closed world-sheet correlators. 
 
 In addition to this, a new subtle zero momentum-effect showed up \cite{Sen:2020cef} in the massless sector, having  to do with the auxiliary massless field $c_0\ket0$. This state is set to zero by the standard Siegel gauge condition $b_0\ket\psi=0$ but since it is in the kernel of $L_0$ it should remain as an external state after the standard integration-out performed by the open string propagator $b_0/L_0(1-P_0)$  (where $P_0$ is the projector on the kernel of $L_0$). The integration-out of this extra mode needs a different propagator, defined in the kernel of $L_0$, which apparently gives new contributions to the string amplitudes which are not accounted for by the standard Polyakov path integral\footnote{The structure of the needed propagator has also been independently analyzed in \cite{Erbin:2020eyc}.}.  The auxiliary zero mode $c_0\ket0$  has also been crucial to obtain the correct normalization $\cal N$ in front of the D-instanton amplitude, which requires a well defined calculation of $e^{\rm annulus}$.  Indeed a correct definition of the annulus amplitude with boundaries on the D-instanton receives a non-trivial contribution from the state $c_0\ket0$ running in the loop, since this field cannot be set to zero by the Siegel gauge condition, which is inconsistent in the zero mode sector. This world-sheet computation has been originally performed in the $c=1$ string \cite{Sen:2021qdk}, finding agreement with the matrix model and in Type IIB \cite{Sen:2021tpp}, providing a  check of S-duality \cite{Green:1997tv} and later generalized to several other backgrounds receiving contributions from D-brane instantons \cite{Agmon:2022vdj,  Alexandrov:2021shf, Alexandrov:2021dyl, Eniceicu:2022nay, Alexandrov:2022mmy}.

\section{Classical Solutions and Background Independence}\label{chap:4}

So far the discussion has hinged on how the SFT framework is useful to give a correct construction of string perturbation theory when infrared divergences are present. 
However this is not the original motivation for  SFT. The idea behind SFT is that it should furnish, through its target space path integral, a non-perturbative definition of the full string theory. In this respect one should expect that non-perturbative contributions to observables should arise from the SFT path integral as saddle points of the  SFT action. Just as it happens with Yang-Mills instantons, this non-perturbative exploration of the string theory landscape has to start with the search for classical solutions.  Ideally, starting from a consistent classical closed string field theory,  all other exact string backgrounds should arise as classical solutions where the various closed string fields acquire VEV's describing a new vacuum state, possibly `far' from the initial one.  As of today, however, this remains in the status of wishful thinking except for some example of perturbative marginal and nearly marginal deformations \cite{Michishita:2006dr, Mukherji:1991tb}. There has also been some numerical searches for zero-momentum tachyon-condensation \cite{Belopolsky:1994sk, Yang:2005rx} and the twisted tachyon condensation associated to the decay of unstable orbifolds \cite{Okawa:2004rh} but the results were not conclusive. The main reason why it is so difficult to search for solutions of closed string field theory  is that the equation of motion is non-polynomial
\begin{align}
Q\Phi+\sum_{k=1}^\infty \frac1{k!}\,l_k(\Phi^{\wedge k})=0
\end{align}
and, as an extra layer of complication, the products $l_k$'s are  not  even known in a concrete enough form, if not for the first few perturbative levels.  Despite the active  on-going research on the characterization of the closed string field theory vertices $l_k$, it  seems hard to gain in the near future a reasonable control on the landscape of closed string field theory. 

 \subsection{D-branes as classical solutions in OSFT}
The situation is definitely better in OSFT. Starting from the simpler model of the bosonic string in the Witten cubic formulation \cite{Witten:1985cc},  the equations of motion are in this case quadratic
\begin{align}
Q\Psi+\Psi*\Psi=0,
\end{align}
where Witten star product $*$ is a precise realization of the open string product $m_2$ \eqref{m_n} which is associative. 
A solution $\Psi_*$ to this equation should represent a new background for the open string theory, in other words a new D-brane system.  An important question is how much of the D-brane landscape is reachable from a starting BCFT$_0$ where the OSFT is initially defined. In critical bosonic string theory all D-brane boundary conditions contain the identity representation in their open string spectrum, from which it is possible to construct the zero-momentum tachyon field
\begin{align}
\ket T=t\,c_1\ket0_{SL(2.\mathbb{R})}.
\end{align}
The first important solution of bosonic OSFT is the so-called tachyon vacuum which represents a (non-perturbative) state where the zero momentum tachyon field has acquired a vev $t_{\rm tv}$. This solution $\Psi_{\rm tv}$ is present on any D-brane system of critical bosonic string theory and represents a configuration where the D-brane has fully decayed leaving behind the pure reference closed string background. This solution can be written in the universal sector containing only $bc$ oscillator excitations and matter Virasoro operators.

This solution was first found in level truncation by Sen and Zwiebach \cite{Sen:1999nx}. Later in 2005 an analytic solution was constructed by Schnabl \cite{Schnabl:2005gv} where also the on-shell value of the classical action was computed and shown to match with the mass of the initial D-brane
\begin{align}
S[\Psi_{\rm tv}]=\frac1{2\pi^2 g_s}Z_{\rm disk}^{\rm BCFT_0}.
\end{align}

This confirmed the expectation that the tachyon vacuum solution indeed represents a configuration with no D-branes. A further confirmation of this hypothesis came soon after \cite{Ellwood:2006ba} with a constructive proof that the kinetic operator around the tachyon vacuum
\begin{align}
Q_{\rm tv}\equiv Q+[\Psi_{\rm tv},\cdot],
\end{align}
supports no cohomology, {\it i.e.} no physical open string states. This has been shown by explicitly constructing  a contracting-homotopy string field $A$ obeying
\begin{align}
Q_{\rm tv}A=1.
\end{align}
This guarantees that any $Q_{\rm tv}$-closed state $\Psi$ is also exact since it can be written as $\Psi=Q_{\rm tv}(A*\Psi)$.

Although the tachyon vacuum supports no perturbative physical excitations, a rich structure of non-perturbative classical solutions is nevertheless expected. These solutions should indeed represent the other possible D-branes that can exist in the given closed string background.  These solutions have been searched (and often found) for a long time, initially as marginal boundary deformations of the perturbative vacuum \cite{Schnabl:2007az, Kiermaier:2007ba, Fuchs:2007yy, Kiermaier:2007vu, Kiermaier:2010cf, Maccaferri:2014cpa} then, more ambitiously,  as IR fixed points of relevant boundary deformations \cite{Ellwood:2009zf, Bonora:2010hi}. However some intrinsic singularities in the equation of motion have been found \cite{Erler:2011tc}. Similar solutions with similar problems with the equation of motion have been proposed for describing multi-brane configurations \cite{Murata:2011ep, Hata:2011ke}, i.e states representing multiple copies of the initial perturbative vacuum. The possible existence of these solutions is in a sense surprising as they should correspond to states with higher energy with respect to the initial perturbative vacuum, which corresponds to `climb up' the boundary world-sheet RG-flow,  which seems  forbidden in a Wilsonian sense.  During the same time however, evidence of positive-energy solutions has emerged in level truncation \cite{Kudrna:2014rya}. This was an indication that OSFT solutions don't necessarily correspond to boundary RG flows from a UV to a IR fixed point but that, more generally, they simply correspond to a change in the world-sheet conformal boundary conditions, irrespectively of whether the disk partition function decreases or increases. This possibility was finally made fully concrete in \cite{Erler:2014eqa, Erler:2019fye}, where a  construction of any BCFT$_*$ as a solution to the OSFT on any other BCFT$_0$ has been given in terms of boundary-condition-changing (bcc) operators, thanks to a rather non-trivial generalization of a previous construction for marginal deformations \cite{Kiermaier:2010cf}. 
This `interpolating' solution is rather easy to write down as
\begin{align}
\Psi_*=\Psi_{\rm tv}^{(0)}-\Sigma\,\Psi_{\rm tv}^{(*)}\,\overline\Sigma.\label{EM}
\end{align}
In this expression $\Psi_{\rm tv}^{(0)}$ and $\Psi_{\rm tv}^{(*)}$ are two tachyon vacuum solutions of the OSFT's defined respectively in the reference BCFT$_0$ and in the target BCFT$_*$. The string fields $\Sigma$ and $\overline \Sigma$ are constructed with bcc operators from BCFT$_0$ to BCFT$_*$ and viceversa, so that the second term in \eqref{EM} is a state in BCFT$_0$, where the equation of motion is defined. In order for \eqref{EM} to be a solution $(\Sigma,\overline\Sigma)$ have to satisfy the properties
\begin{align}
Q_{\rm tv}\Sigma&=0\\
Q_{\rm tv}\overline\Sigma&=0\\
\overline\Sigma\,\Sigma&=1.
\end{align}
In particular in \cite{Erler:2019fye} it has been shown that any pair of bcc operators containing the identity representation of BCFT$_*$ in their OPE
\begin{align}
\bar \sigma(x) \sigma(0)= x^{-2 h} \boldsymbol{1}_{\textrm{BCFT}_*}+\textrm{less singular},
\end{align}
can be used to explicitly construct $(\Sigma,\overline\Sigma)$. The action evaluated on the solution correctly reproduces the shift in the disk partition function between the starting and the final open string background
\begin{align}
S[\Psi_*]=\frac1{2\pi^2g_s}\left(Z_{\rm disk}^{\rm BCFT_0}-Z_{\rm disk}^{\rm BCFT_*}\right).
\end{align}
Even more interestingly, the solution provides a rather neat example of background independence. Indeed, thanks to the properties obeyed by $(\Sigma,\overline\Sigma)$, one can consider the field redefinition\footnote{Technically, this is a field redefinition only after the integration out of a trivial non-dynamical sector.}
\begin{align}
f:{\cal H}_{\textrm{BCFT}_*}&\to{\cal H}_{\textrm{BCFT}_0}\\
f(\phi)&\coloneqq \Sigma \phi\overline\Sigma,
\end{align}
and explicitly verify that
\begin{align}
S[\Psi_*+\Sigma \phi\overline\Sigma]=\frac1{2\pi^2g_s}\left(Z_{\rm disk}^{\rm BCFT_0}-Z_{\rm disk}^{\rm BCFT_*}\right)+S^{(*)}[\phi],
\end{align}
where $S^{(*)}[\phi]$ is the OSFT action directly defined on the new background BCFT$_*$.

Some of these results generalize to the superstring \cite{Erler:2019nmz}. In this case however the most useful open superstring field theory is Berkovits's  theory in the LHS, which does not require any insertion of PCO's.
If the theory is formulated on a non-BPS D-brane, or on a brane/anti-brane pair, then  it is possible to construct a tachyon vacuum solution supporting no physical open string states and having the correct energy density \cite{Erler:2013wda}.
The possibility of constructing the interpolating solution is less clear in this case since its construction hinges on the existence of the tachyon vacuum for both the initial and the final background, but this seems to exist only for unstable backgrounds. For example, a typical solution that is expected to be found is the formation  of  BPS $D(p-2)$ branes as `tachyon vortices' \cite{Sen:1998tt} of a $Dp$-$\overline{Dp}$ pair. In this case, the RR charge of the final background should be captured by a topological charge of the solution, associated with the winding of the two-dimensional tachyon field around the $U(1)$ vacuum manifold of the tachyon potential of the $Dp$-$\overline{Dp}$ system. However, despite some  preliminary discussion of a possible emerging topology in the space of superstring field theory solutions \cite{Erler:2013wda}, the  emergence of such topological charges is still mysterious. This suggests that there is still a lot to understand concerning the space of solutions of the open superstring. 

\subsection{D-branes and closed strings}

Working in the framework of pure OSFT does not allow to change the closed string background and the notion of background independence is thus limited. In more interesting situations  D-branes can have rich interactions with closed strings in the bulk. This has been analyzed in the framework of open-closed SFT in a couple of complementary approaches. In the first approach one can ask the question of how a D-brane system is modified by a change in the closed string background \cite{Zwiebach:1997fe}. At the classical level, and ignoring the D-brane backreaction on the closed string background, this can be answered by truncating the open-closed SFT to interaction vertices which only include spheres and disks. In this truncation it is possible to first change the closed string background with a closed SFT solution $\Phi_*$ and then to perform a vacuum shift in the open string background $\Psi_{\rm v}$. The natural observable that is associated to this process is the vacuum energy $\Lambda$ that is generated by the the closed SFT solution $\Phi_*$ and the open string vacuum shift $\Psi_{\rm v}$. This observable has been shown to compute the shift in the total world-sheet disk partition function between the initial D-brane in the initial closed string background and the final D-brane in the deformed closed string background \cite{Maccaferri:2022yzy}
\begin{align}
\Lambda=\frac1{2\pi^2}\left(\frac1{g_s^{(0)}}Z_{\rm disk}^{\rm BCFT_0}-\frac1{g_s^{(*)}}Z_{\rm disk}^{\rm BCFT_*}\right).
\end{align}
Notice that in this case (where the closed string background changes) also the string coupling constant will change. This observable is useful because it can be used to extract physical information on the closed string background described by $\Phi_*$. Analogous purely closed string quantities are indeed known to formally vanish \cite{Erler:2022agw}.

Perhaps more interesting and challenging is the opposite question about how to characterize the change in the closed string background due to the presence of a D-brane system, in other words the D-brane's backreaction. Again this is a question that can be  answered in open-closed SFT by considering a large  number $N$ of initial identical D-branes. In this case  the quantum behaviour of open-strings  cannot be ignored,  but the large $N$ limit allows to keep the closed string sector classical and to only consider the planar sector of quantum open strings \cite{Maccaferri:2023gof}. In this case, the algebraic structure of the open-closed SFT in the large $N$ limit strongly suggests that a new classical closed string field theory without D-branes  can be obtained,  by first  integrating out the open string sector and then by stabilizing the obtained closed string field theory to eliminate the tadpole created by the world-sheet boundaries. This mechanism precisely fits the known examples of exact open-closed dualities \cite{Gopakumar:1998ki, Gaiotto:2003yb} and further explorations will hopefully shed more light on this approach towards a microscopic understanding of the gauge/gravity duality \footnote{See also \cite{Okawa:2020llq} for related ideas and approaches to open/closed duality in SFT.}.

\section{Conclusions }\label{chap:5}

In the last decade SFT has emerged as the most appropriate approach to deal with string perturbation theory. Much progress has been achieved on the front of background independence, through the understanding of classical solutions. Much more remains to be done and understood at various levels. Some of the main open problems are listed below.
\begin{itemize}
\item Explicit construction of interaction vertices both at classical and quantum level for closed string field theory.
\item Explicit construction of interaction vertices for the superstring at the quantum level, with a correct distribution of PCO's such that spurious singularities are avoided.
\item Construction of solution to closed string field theory. This can hopefully be achieved either by explicit constructions of interaction vertices, or by possible reformulations of closed SFT with auxiliary fields to make it cubic or polynomial.
\item Construction of vacuum-shift solutions canceling large NS-NS tadpoles which cannot be handled perturbatively.
\item Construction of solutions to open superstring field theory.
\item Characterization of the various aspects of open-closed duality in a genuine SFT framework with the aim of penetrating the microscopic origin of holography. This includes a better understanding of the status of OSFT as a possible quantum theory, possibly depending on the given string background.
\item Eventually, classical solutions should be used to capture the non-perturbative content of String Theory, but how to do this concretely is still rather elusive. 
\end{itemize}
There are other interesting aspects which, because of the limited  space, it has not been possible to review.  These include the use of SFT to explore the space of two-dimensional (B)CFT's via the correspondence between (O)SFT solutions and (B)CFT's \cite{Kudrna:2012re, Kojita:2016jwe, Scheinpflug:2023osi, Scheinpflug:2023lfn}  and the related progress in the level truncation techniques to systematical scan the space of OSFT solutions \cite{Kudrna:2014rya, Kudrna:2016ack, Kudrna:2019xnw, Kudrna:2021rzd}. 

The whole subject of SFT is steadily progressing and new developments on various fronts are to be expected in the near future.

\section*{Acknowledgments}
The author acknowledges useful discussions at various times with Loriano Bonora, Harold Erbin, Ted Erler, Matej Kudrna,  Raghu Mahajan, Yuji Okawa, Ivo Sachs, Martin Schnabl, Ashoke Sen,  Jakub Vosmera, Xi Yin and  Barton Zwiebach.
The author thanks A. Sen for comments on the draft.
This work is partially supported by the MUR PRIN contract 2020KR4KN2 “String Theory as a bridge between Gauge Theories and Quantum Gravity” and by the INFN project STeFI “String Theory and Fundamental Interactions”.

\medskip

\bibliographystyle{ieeetr} 
\bibliography{refs} 

%
%
%
\end{document}